\documentstyle[aps,prl]{revtex}
\begin{document}
\draft

\title{Gauge Interactions and Bosonized Fermion Liquids}
\author{H.-J. Kwon, A. Houghton, and J. B. Marston}
\address{Department of Physics, Brown University, Providence, RI 02912-1843}
\date{January 20, 1994}
\maketitle

\begin{abstract}
We investigate fermion liquids interacting with longitudinal and transverse
abelian gauge fields via bosonization.   In two spatial dimensions we obtain
the fermion propagator for the specific case of a Coulomb plus Chern-Simons
gauge action.  We discuss the relevance of this result to the
Halperin-Lee-Read theory of the $\nu =1/2$ Landau level and demonstrate how
Kohn's theorem is satisfied.
\end{abstract}

\pacs{05.30.Fk, 11.40.-q, 72.20.My, 73.40.Hm}

cond-mat/9401041

\medskip
The many body problem of fermions coupled to gauge fields is central to
the physics of strongly correlated electronic systems.
Gauge fields play an important role in the physics of the half-filled
Landau level\cite{Halperin} and may be important in
the cuprate superconductors\cite{Gauge,Lee}.
The effects of transverse gauge fields,
which are not screened, are particularly interesting as perturbative
expansions in the coupling constant break down\cite{Blok}.
Renormalization group analysis suggests that a novel
fixed point exists\cite{Gan}.
Bosonization in dimensions greater than one\cite{Haldane,Tony,HKM,HKMS} is
a powerful tool for understanding the effects of gauge fields on fermion
liquids as the realm in which it is applicable, for low-energy excitations
near the Fermi surface, is precisely where these interactions are most
important.  Bosonization does not rely upon a Fermi liquid
form for the quasiparticle propagator; consequently non-trivial fixed
points are accessible.  In an appropriate limit, bosonization reduces
complicated many-body Hamiltonians to quadratic forms.  In this
paper we consider the general problem of longitudinal and transverse gauge
fields coupled to a Fermi gas in two dimensions.  Specializing to
the Chern-Simons theory studied by Halperin, Lee, and Read\cite{Halperin}
(HLR) we show that a novel
non-Fermi-liquid fixed point controls the low energy physics.
Furthermore, by including the Landau parameter $f_1$
we automatically recover Kohn's theorem\cite{Kohn}: the bare
cyclotron frequency, not a renormalized value, appears in the collective mode
spectrum.

For simplicity, we confine our attention to the case of spinless fermions
and spherical Fermi surfaces and set $\hbar = 1$.
We begin with the bare Hamiltonian for fermions
interacting with non-compact longitudinal and transverse gauge
fields $A^{\mu}({\bf q})$.  In the Coulomb gauge, ${\bf \nabla \cdot A} = 0$
it is:
\begin{equation}
H = \int d^Dx ~{1\over 2m}c^{\dag}({\bf x })\big{[}(-i{\bf \nabla -A})^2
- A_0 \big{]}c({\bf x}) +{{1}\over{2}}~
\int d^Dx~ d^Dy~ V({\bf x-y})~c^{\dag}({\bf x})~ c^{\dag}({\bf y})~
c({\bf y})~ c({\bf x})
\label{bare}
\end{equation}
where $\epsilon_{\bf k} \equiv {\bf k}^2/2m$ for the case of a Fermi gas
and $V({\bf x-y})$ represents Coulomb or other two-body interactions.
Next we integrate out the
high-energy Fermi degrees of freedom with the use of the renormalization
group\cite{Shankar}.  The resulting low-energy effective theory is
expressed in terms of quasiparticles $\psi_{\bf k} $ which obey
canonical anticommutation relations and which are
related to the bare fermion operators via the wavefunction renormalization
factor $Z_{\bf k}$:
\begin{equation}
\psi _{\bf k} = Z^{-1/2}_{\bf k} ~ c_{\bf k}
\end{equation}
for momenta $\bf k$ restricted to a narrow shell of thickness $\lambda$
around the Fermi surface:
$k_F - \lambda/2 < |{\bf k}| < k_F + \lambda/2$.
The factor $Z_{\bf k}$ rescales the discontinuity in the quasiparticle
occupancy at the Fermi surface back to one.  The operation of integrating
out the high energy fields can be written explicitly by introducing the
partition function:
\begin{equation}
Z = \int {\cal D}A^\mu ~{\cal D}\psi ~{\cal D}\not\!\psi ~e^{-i~S[A,~ \psi,~
\not\!\psi]}
\end{equation}
Here, $\not\!\psi $ are the high-energy Fermi fields that are integrated
out first.  Introducing the D+1-vector notation $q \equiv (\omega, {\bf q})$,
the resulting low-energy effective action may be written as:
\begin{equation}
S[A, \psi] = \int d^Dx~dt ~\big{[}\psi^{*}({\bf x })(i\partial _t +A_0)
\psi({\bf x})
+{1\over 2m^*}\psi^{*}({\bf x})({\bf \nabla}-i~{\bf A})^2 \psi({\bf x})\big{]}
+ \sum_{\bf q}~ \int {{d\omega}\over{2 \pi}}~
\not\!\Pi_{\mu \nu}(q)~ A^\nu(q)~ A^\mu(-q)~ .
\label{FL}
\end{equation}
In addition there are two-body fermion interactions which in general include
both the long range Coulomb interaction and short range Fermi liquid type
interactions which we include later.
There are also irrelevent operators consisting of higher powers and/or
derivatives
of the gauge and Fermi degrees of freedom which may safely be neglected in the
low-energy limit\cite{Gan}.
The quadratic $\not\!\Pi_{\mu \nu}$ term is generated by integrating out the
$\not\!\psi$ fields deep inside the Fermi sea.  It contributes to the
total diamagnetic term of
$-(\rho_f / 2 m^*) {\bf A}(q) {\bf \cdot} {\bf A}(-q)$.

We now review the main aspects of bosonization and refer the reader to
previous papers\cite{Tony,HKM,HKMS} for more details.  In $D$-spatial
dimensions, the Fermi quasiparticle
fields $\psi$ may be expressed\cite{Tony} in terms of the charge boson fields
$\phi$ as:
\begin{equation}
\psi({\bf S; x}, t) = {1\over\sqrt{V}}~ \sqrt{{\Omega}\over{a}}
e^{i{\bf k}_{\bf S}{\bf \cdot x}}
\exp \{i{\sqrt{4\pi}\over \Omega} \phi({\bf S; x}, t)\}~ O({\bf S}),
\label{bosonization}
\end{equation}
where $\bf S$ labels the patch
[${\bf S} \equiv (\theta, \phi)$ in three dimensions]
on the Fermi surface with momentum ${\bf k_S}$.  $V$ is the volume of the
system which equals $L^D$ in D-dimensions; the factor of $\sqrt{V}$
is introduced to maintain canonical fermion anticommutation relations.
Both $\psi$ and $\phi$ fields are defined inside a squat
box centered on $\bf S$ with height $\lambda$ in the radial (energy) direction
and area $\Lambda^{D-1}$ along the Fermi surface.
These two scales must be small in the
following sense: $k_F >> \Lambda >> \lambda$.  We satisfy these limits
by setting $\lambda \equiv k_F/N$ and $\Lambda \equiv k_F/N^\alpha$ where
$0 < \alpha < 1$ and $N \rightarrow \infty$.
The quantity $a$ in the bosonization formula Eq. (\ref{bosonization})
is a real-space cutoff given by $a \equiv 1/\lambda$.  Here
$\Omega \equiv \Lambda^{D-1} (L/2 \pi)^D$ equals the number of states
in the squat
box divided by $\lambda$.  Finally, $O(S)$ is an ordering operator
introduced\cite{Tony,Luther}
to maintain Fermi statistics in the angular direction along
the Fermi surface.  (Anticommuting statistics are obeyed automatically
in the radial direction.)
The charge current in a given Fermi surface patch $\bf S$
is defined by:
\begin{equation}
J({\bf S; q}) \equiv \sum_{\bf k} \theta({\bf S; k + q})~
\theta({\bf S; k})~ \{ \psi^{\dagger}_{\bf k + q}~
\psi_{\bf k} - \delta^3_{\bf q, 0}~ n_{\bf k} \}\ .
\label{curk}
\end{equation}
Here $\theta({\bf S; k}) = 1$ if ${\bf k}$ lies inside a squat box
of dimensions $\lambda \times \Lambda^{D-1}$ centered
at $\bf S$ and equals zero
otherwise.  The charge currents obey the abelian Kac-Moody algebra\cite{Tony}.

We now rewrite the effective action Eq.(\ref{FL}) in terms of charge currents
$J$:
\begin{eqnarray}
S[A,a] &=& \sum_{\bf S} \sum_{\bf q, q\cdot \hat{n}_S >0}
\int {d\omega \over 2\pi}
{}~(\omega - v_F^*~ {\bf q\cdot \hat{n}_S})~a^{*}({\bf S}; q)~a({\bf S}; q)
\nonumber \\
&+& {1\over V} \sum_{\bf q }\int {d\omega \over 2\pi }
\big{\{} \sum_{\bf S} J({\bf S}; q)[ A_0(-q) +
{\bf v_S^*\cdot A}(-q)] -{\rho_f\over 2 m^*} {\bf A}(q)\cdot {\bf A}(-q)
\big{\}} ~.
\label{action}
\end{eqnarray}
Here ${\bf v_S^* \equiv k_S} / m^*$ is the renormalized Fermi velocity vector
at patch $\bf S$ and
the charge currents are related to the canonical boson operators $a$ and
$a^\dagger$ by:
\begin{equation}
J({\bf S; q}) = \sqrt{ \Omega~ |{\bf \hat{n}_S \cdot q}|}~
[ a({\bf S; q})~ \theta({\bf \hat{n}_S \cdot q}) + a^\dagger({\bf S; -q})~
\theta(-{\bf \hat{n}_S \cdot q})]\ . \label{canon}
\end{equation}
Also $\theta(x)=1$ if $x>0$ and is zero otherwise and
$\rho_f$ is the full fermion number density.
Exchange scattering, which involves high momentum gauge fields, has not been
included in the action, as it is not singular in the low-momentum limit.
We also assume that no superconducting or charge-density-wave instabilities
occur.  The bare gauge action, which we have not yet discussed, is in
general a quadratic form:
\begin{equation}
S^0_G[A] = {1\over 2}\int {d^Dq\over (2\pi)^D}\int
{d\omega \over 2\pi }~ K^0_{\mu \nu}(q)~ A^\mu(q)~ A^\nu(-q)~.
\label{SG0}
\end{equation}
Below we will specialize to the particular case of the Chern-Simons action.

We proceed to evaluate the exact boson correlation function.
Rather than integrate out the gauge fields to obtain an effective
current-current interaction, we instead introduce a generating functional
for the boson correlation function and first
integrate out the boson fields $a$ and
$a^*$.  In this way we avoid the intermediate step of summing an infinite
perturbation series in powers of the effective interaction to obtain the
boson propagator.  We construct the generating functional by coupling
fields $\xi, \xi^*$ to the boson fields $a^{*}, a$.
Then, by completing the square, we integrate out the boson fields to obtain the
effective action of the gauge fields.  Since the bosons are linearly coupled
to the gauge fields the resulting effective action is quadratic in the
gauge fields $A^\mu$.  The integral over the $A^\mu$ fields can be carried out
to yield the generating functional explicitly as a function of $\xi, \xi^*$.
\begin{eqnarray}
Z[\xi, \xi^*] &=& \int {\cal D}A^\mu ~{\cal D}a~{\cal D}a^{*}~
\exp \{i(S[A,a]+S^0_G[A])\}~ \exp \big{\{}
i\sum_{\bf S}\sum_{\bf q, q\cdot \hat{n}_S>0}
\int {d\omega \over 2\pi }~[\xi ({\bf S}; q)~ a^{*}({\bf S}; q)
+ \xi^*({\bf S}; q)~a({\bf S}; q)] \big{\}} \nonumber \\
&=& {\cal N}\int {\cal D}A^\mu ~\exp\{i(S_{eff}[A]+S^0_G[A])\}
\nonumber \\
&\times&
\exp \big{\{} -i\sum_{\bf S}\int{d^Dq\over (2\pi)^D}~\theta({\bf q\cdot
\hat{n_S}})
\int {d\omega \over 2\pi }~{\sqrt{\Omega {\bf q\cdot \hat{n}_S}}
\over \omega -v_F^* {\bf q\cdot \hat{n}_S}}~
[\xi ({\bf S}; q)~ (A_0(-q)+ {\bf v_S^* \cdot A}(-q)) \nonumber \\
&+& \xi^*({\bf S}; q)~(A_0(q) +
{\bf v_S^* \cdot A}(q))] - i \sum_{\bf S}\sum_{\bf q, q \cdot \hat{n}_s>0}
\int {d\omega \over 2\pi }~{1\over \omega -v_F^*~ {\bf q\cdot \hat{n}_S}}~
\xi({\bf S}; q)~ \xi^*({\bf S}; q)\big{\}}~ .
\label{ftnal}
\end{eqnarray}
The boson progator is obtained by differentiating twice:
\begin{equation}
\langle a({\bf S}; q)~ a^{\dagger}({\bf S}; q)\rangle
= -{{\delta^2 \ln Z[\xi, \xi^*]}\over{\delta \xi({\bf S}; q)
{}~\delta \xi^*({\bf S}; q)}}|_{\xi = \xi^* = 0}~.
\label{derivat}
\end{equation}
In Eq.(\ref{ftnal}), the effective action equals:
\begin{equation}
S_G[A] = S_{eff}[A] + S^0_{G}[A] = {1\over 2}\int {d^Dq\over (2\pi)^D}\int
{d\omega \over 2\pi }~ K_{\mu \nu}(q)~ A^\mu(q)~ A^\nu(-q)~,
\label{SG}
\end{equation}
where the integrand is:
\begin{equation}
K_{\mu \nu }(q)~ A^\mu(q)~ A^\nu(-q) = K^0_{\mu \nu}(q)~ A^\mu(q)~ A^\nu(-q)
+ \chi ^0(q)~ A_0(q)~ A_0(-q) + \chi^T(q)~ {\bf A}(q) \cdot {\bf A}(-q)~ .
\label{Kmn}
\end{equation}
Here, $\chi^0$ and $\chi^T$ are the longitudinal and
transverse susceptibilities given in two spatial dimensions by:
\begin{equation}
\chi ^0(q) = N^*(0) \big{[} 1 -{\theta(x^2-1)~|x|
\over \sqrt{x^2-1}}+i{\theta(1-x^2)~|x|\over \sqrt{1-x^2}}\big{]}
\label{chi0}
\end{equation}
and
\begin{equation}
\chi ^T(q) = v_F^{* 2}~ N^*(0) \big{[} -x^2 +\theta(x^2-1)~|x|\sqrt{x^2-1}
+i\theta(1-x^2)~|x|\sqrt{1-x^2} \big{]}
\label{chiT}
\end{equation}
where $x \equiv {\omega \over v_F^*~|{\bf q }|}$ and $N^*(0) = m^*/2\pi$
is the quasiparticle density of states at the Fermi surface.
The boson propagator can now be determined exactly from
Eqs. (\ref{ftnal}, \ref{derivat}, \ref{SG}).  Expressed in terms of the gauge
propagator, $D_{\mu \nu}(q) = \big{[} K(q) \big{]}^{-1}_{\mu \nu}$, it is:
\begin{eqnarray}
\langle a({\bf S}; q)~ a^\dagger({\bf S}; q)\rangle
&=& {i\over \omega - v_F^* {\bf q\cdot \hat{n}_S} +i \eta~ {\rm sgn}(\omega)}
+ i~ {\Lambda^{D-1} \over (2\pi )^D}~ {\bf q\cdot \hat{n}_S}~
{D_{\mu \nu }({\bf q},\omega )~\epsilon ^\mu ({\bf S})
{}~\epsilon ^\nu ({\bf S})\over
[\omega - v_F^* {\bf q\cdot \hat{n}_S} + i \eta ~{\rm sgn}(\omega)]^2} ~ ,
\label{Gaa}
\end{eqnarray}
where $\epsilon ^\mu ({\bf S})$ is the D+1-vector $(1, {\bf v_S^*})$.

Next we turn to the problem of the half-filled Landau level.
The HLR theory\cite{Halperin}
describes this system as a collection of
quasiparticles which obey Fermi statistics.
Each quasiparticle is a composite object consisting of the physical electron
plus a flux tube of quanta $\tilde{\phi}$.  The
attachment of flux tubes to quasiparticles is natural in a Chern-Simons
theory.  The average field strength of the flux tubes is arranged to
cancel out the external magnetic field.
Thus, at the mean-field level, the quasiparticles behave as free fermions in
zero net field.  Consequently, the infinite Landau degeneracy is lifted and
presumably the state is stable against two-body Coulomb interactions.
The important question then is whether gauge fluctuations (or equivalently
quasiparticle density fluctuations) modify
or destroy this mean-field state.  Unlike physical Maxwell electromagnetic
gauge fields, the transverse component of the statistical gauge field is not
suppressed by a numerical factor of the fine structure constant multiplied
by the square of the Fermi velocity to the speed of light ratio.
In the following we assume that despite strong gauge fluctuations
the Fermi surface remains well enough defined for our bosonic construction
to apply and later verify that this assumption is
correct.  Experimental evidence supports the existence of a Fermi
surface\cite{Willett}.  The gauge fields in the HLR theory
are compact and instantons may play an important role in the electron
Green's function\cite{Kim}.  For simplicity we ignore these effects here.
In the Coulomb gauge the HLR effective action is:
\begin{eqnarray}
S &=& \int d^2x~dt~\big{[} \psi ^{\dag}(i~\partial_t +A_0)\psi
+ {1\over 2m^*}\psi ^{\dag}({\bf \nabla }-i~{\bf A})^2\psi
+ {A_0\over 2\pi \tilde{\phi}} ({\bf \nabla \times A}) \big{]} \nonumber \\
&-& {1\over 2}~{1\over (2\pi \tilde{\phi })^2}\int d^2x~d^2y~dt~
{\bf [\nabla_x \times A(x)]~[\nabla_y \times A(y)] }~V({\bf x -y})~ .
\label{SFQH}
\end{eqnarray}
At filling fraction $\nu =1/2$ we set $\tilde{\phi} =2$ to cancel out the
external field.  The last term in Eq. (\ref{SFQH})
is the two-body Coulomb interaction, now written in terms of the gauge field
on making use of the constraint that a flux tube is attached to each
electron\cite{Halperin}.
We should also include Fermi liquid interactions, which if all coefficients
except $f_1$ are set equal to zero, contribute to the action a term of the
form:
\begin{equation}
S_{FL} = -{{f_1}\over{2 V k_F^2}}~ \int {{d\omega}\over{2\pi}}~
\sum_{S,T,{\bf q}}~ J(S; q)~ {\bf k_S \cdot k_T}~ J(T; -q)
\label{f1}
\end{equation}
which is made gauge covariant by the replacement\cite{Simon}
$\bf k = -i \nabla \rightarrow -i \nabla - A$.
Given the action Eq. (\ref{SFQH}) the bare gauge term Eq.(\ref{SG0}) becomes:
\begin{eqnarray}
S_{G0} &=& {i\over 4 \pi \tilde{\phi}}
\int {d^2q\over (2\pi)^2}\int {d\omega \over 2\pi }~
\{ A_0(-q)~ [{\bf q \times A}(q)] - A_0(q)~ [{\bf q \times A}(-q)] \}
\nonumber \\
&-& {1\over 2}~{1\over (2\pi \tilde{\phi})^2} \int {d^2q\over (2\pi)^2}
\int {d\omega \over 2\pi }~{\bf q}^2~ V({\bf q})~
[{\bf A}(q)~{\bf \cdot A}(-q)]~ ,
\label{CS}
\end{eqnarray}
where $V({\bf q})$ is the Fourier transform of $V({\bf x -y})$.
For the Coulomb interaction, $V({\bf q}) = {2\pi e^2\over |{\bf q}|}$,
where $e^2 = \alpha c$ and $\alpha \approx 1/137$
is the fine-structure constant.
In the q-limit of $\omega << |{\bf q}|$ we find the gauge propagator to
be given by:
\begin{equation}
D_{\mu \nu }(q)~\epsilon^\mu ({\bf S}) ~\epsilon^\nu ({\bf S})
\simeq {{|{\bf q\times \hat{n}_S}|^2}\over{{\bf q}^2}}
{{v_F^{* 2}~ N^*(0)}\over{i {{\rho_f |\omega|}\over{\pi v_F^*|{\bf q}|}}
- {{{\bf q}^2}\over{(2\pi \tilde{\phi})^2}} [1 + N^*(0)~ V({\bf q})]}}~ .
\label{damp}
\end{equation}
In this limit the boson self energy has an imaginary part which reflects
quasiparticle damping.
For the Coulomb interaction the pole is located
at $\omega \sim i~e^2 {\bf q}^2/k_F$.
If, on the other hand, $V({\bf q})$ is a short-range interaction,
the pole is at $\omega \sim i~v_F^*~ |{\bf q}|^3/k_F^2$ in agreement
with the RPA result\cite{Halperin,Lee}.  In the opposite $\omega$-limit of
$\omega >> |{\bf q}|$ we obtain instead:
\begin{equation}
D_{\mu \nu }(q)~\epsilon^\mu ({\bf S}) ~\epsilon^\nu ({\bf S})
\simeq {1\over 2}~
{\omega ^2\over {\bf q}^2}~ {{v_F^{* 2}~ N^*(0)~ (2\pi \tilde{\phi})^2}\over
{\omega ^2 -  (2 \pi \tilde{\phi}~ \rho_f)^2/m^{* 2}}} ~ .
\label{gap}
\end{equation}
The poles in the $\omega$-limit which characterize
collective excitations\cite{HKM} are found at
$|\omega | = {2\pi \rho_f\tilde{\phi}\over m^*}$,
the cyclotron frequency for a free particle of mass $m^*$
in an external magnetic field of $B = {2\pi \rho_f c \tilde{\phi}\over e}$.
As it stands this result violates Kohn's
theorem\cite{Kohn}.  The correct result is obtained on including the Fermi
liquid interaction, Eq. (\ref{f1}),
in the fully quantum mechanical bosonic theory.
The poles are then shifted to $|\omega| = 2\pi \rho_f
({1\over m^*}+{f_1\over 4\pi }) \tilde{\phi}
= {2\pi \rho_f \tilde{\phi}\over m}$ in accord with Kohn's theorem.
The details of the proof of this result will be given elsewhere\cite{next}.
The form of Eq.(\ref{gap}) is similar to that found for the
super long range interaction of Bares and Wen\cite{Wen} and therefore
can be expected to make a small contribution to the anomalous exponent
appearing in the real space fermion Green's function\cite{HKMS}.
We can estimate from Eq. (\ref{damp}) the fermion
quasiparticle self-energy\cite{HKM,HKMS}
by expanding the fermion Green's function
in powers of $D_{\mu \nu}$; however it should be noted that there is no obvious
small parameter to justify the expansion.  Given this caveat,
to first order in the case of a short-range interaction we obtain
${\rm Im}~\Sigma_f \sim |\omega|^{2/3}$ and for the Coulomb interaction
${\rm Re}~\Sigma_f \sim \omega \ln |\omega|$
in agreement with Ref.\cite{Halperin}.

As the expansion in powers of $D_{\mu \nu}$ is unreliable, it is
more useful to determine
the quasiparticle propagator directly in real space and time.
We focus on the physical case of the Coulomb interaction.
Given the bosonization formula Eq.(\ref{bosonization}), the real
space fermion Green's function is obtained by computing the Fourier transform
of
the boson correlation function:
\begin{eqnarray}
\ln G_f({\bf S;x},t) &=& \ln ({a\over {\bf x\cdot \hat{n}_S}-v_F^*~ t})
\nonumber \\
&+& i\int {d^2q\over (2\pi)^2}\int {d\omega \over 2\pi }~
[e^{i({\bf q\cdot x}-\omega t)}-1]~
D_{\mu \nu }({\bf q},\omega )~\epsilon ^\mu ({\bf S})
{}~\epsilon ^\nu ({\bf S})
{}~[\omega -v_F^*~ {\bf q\cdot \hat{n}_S}+i\eta ~{\rm sgn}(\omega )]^{-2} ~ .
\label{fermi}
\end{eqnarray}
To further simplify the calculation we consider only the equal time
propagator by setting $t=0$.  In this limit the integrals appearing in
Eq. (\ref{fermi}) can be evaluated and we obtain:
\begin{equation}
G_f({\bf S;x}) \sim {1\over{\bf \hat{n}_S\cdot x}}~
({{c~ \alpha~ \Lambda ^2}\over
{v_F^*~ k_F~ \tilde{\phi}^2}}~ |{\bf \hat{n}_S\cdot x}|)^{-\zeta}~ ,
\end{equation}
where $\zeta = {{(2\pi \tilde{\phi})^2}\over{4\pi \alpha}}
({{v_F^*}\over{c}})$.
The quasiparticle pole has been destroyed by the transverse gauge interactions.
Apparently
the half-filled Landau level system is controlled by a novel non-Fermi liquid
fixed point.
Despite the fact that the Green's function has a Luttinger liquid form,
the quasiparticle occupancy changes abruptly at the Fermi surface as the
momentum scale for the change is set by
${{c~ \alpha~ \Lambda^2}\over{v_F^*~ k_F~ \tilde{\phi}^2}} \rightarrow 0$
as $\Lambda \rightarrow 0$.  Evidently the fixed
point is more like a marginal Fermi liquid than a Luttinger liquid.
The existence of a
discontinuity agrees with experimental observation\cite{Willett}.
It is also consistent with
the assumed Kac-Moody current algebra, as the derivation of the algebra itself
relies upon the existence of a discontinuity in the quasiparticle
occupancy\cite{Tony}.  The novel fixed point therefore is accessible via
bosonization.

Bosonization of fermion liquids in two or three spatial dimensions coupled
to gauge fields yields non-perturbative insight into strongly correlated
systems.  The generating functional approach is a convenient way to compute
the correlation function of the interacting bosons; in fact
previous results obtained by an exact resummation of a perturbative expansion
for the cases of a
single Fermi liquid interaction parameter $f_0$\cite{HKM} and a pure Coulomb
interaction\cite{HKMS} are reproduced directly by the longitudinal
part of Eq. (\ref{Gaa}).
By applying the method to the HLR theory of the half-filled
Landau level we identified the novel non-Fermi liquid fixed point that
governs the low-energy physics.  Kohn's theorem is satisfied automatically.

H.-J.K. thanks Youngjai Kiem and J.B.M. thanks P. Nelson and R. Shankar
for helpful discussions.
This research was supported by the National Science Foundation
through grants DMR-9008239 (A.H.) and DMR-9357613 (J.B.M.).


\begin{references}
\bibitem{Halperin}B. I. Halperin, P. A. Lee and N. Read, {\it Phys. Rev. B}
{\bf 47}, 7312 (1993).
\bibitem{Gauge}See, for example,
G. Baskaran and P. W. Anderson, {\it Phys. Rev. B} {\bf 37},
580 (1988); I. Affleck and J. B. Marston, {\it Phys. Rev. B} {\bf 37}, 3774
(1988); L. B. Ioffe and A. I. Larkin, {\it Phys. Rev. B} {\bf 39},
8988 (1989).
\bibitem{Lee}P. A. Lee and N. Nagaosa, {\it Phys. Rev. B} {\bf 46},
5621 (1992).
\bibitem{Blok}B. Blok and H. Monien, {\it Phys. Rev. B} {\bf 47}, 3454 (1993).
\bibitem{Gan}J. Gan and E. Wong, {\it Phys. Rev. Lett.} {\bf 71}, 4226 (1993);
Chetan Nayak and Frank Wilczek, ``Non-Fermi Liquid Fixed Point
in 2 + 1 Dimensions,'' Princeton University preprint (cond-mat/9312086).
\bibitem{Haldane}F. D. M. Haldane, ``Luttinger's Theorem and Bosonization
of the Fermi Surface,'' Proceedings of the International School of Physics
``Enrico Fermi,'' Varenna 1992 (R. Schrieffer and R. A. Broglia, eds.).
\bibitem{Tony}A. Houghton and J. B. Marston, {\it Phys. Rev. B} {\bf 48},
7790 (1993).
\bibitem{HKM}A. Houghton, H.-J. Kwon, and J. B. Marston, ``On the Stability
and Single-Particle Properties of Bosonized Fermi Liquids,'' Brown University
preprint (cond-mat/9310043).
\bibitem{HKMS}A. Houghton, H.-J. Kwon, J. B. Marston and R. Shankar,
``Coulomb Interaction and the Fermi Liquid State: Solution by
Bosonization,'' Brown University preprint (cond-mat/9312067).
\bibitem{Kohn}W. Kohn, {\it Phys. Rev.} {\bf 123}, 1242 (1961).
\bibitem{Shankar}R. Shankar, {\it Physica} A{\bf 177}, 530 (1991) and
to appear in {\it Rev. Mod. Phys.}
\bibitem{Luther}A. Luther, {\it Phys. Rev. B}, {\bf 19}, 320 (1979).
\bibitem{Willett}R. L. Willett, R. R. Ruel, K. W. West, and L. N. Pfeiffer,
{\it Phys. Rev. Lett.} {\bf 71}, 3846 (1993)
\bibitem{Kim}Yong Baek Kim and Xiao-Gang Wen, ``Instantons and the spectral
function of electrons in the half-filled Landau level,'' MIT preprint
(cond-mat/9401032).
\bibitem{Simon}Steven H. Simon and Bertrand I. Halperin, ``Finite-Wavevector
Electromagnetic Response of Fractional Quantized Hall States, ''
Harvard University preprint (1993), submitted to {\it Phys. Rev. B}.
\bibitem{next}H.-J. Kwon, A. Houghton, and J. B. Marston, to be published.
\bibitem{Wen}P. Bares and X. G. Wen, {\it Phys. Rev. B} {\bf 48}, 8636 (1993).
\end{references}
\end{document}